# Minkowski Functionals in Cosmology[*]

Jens Schmalzing, Martin Kerscher, Thomas Buchert

Theoretische Physik, Ludwig–Maximilians–Universität

Theresienstr. 37, D–80333 München, Germany

**Abstract.** Minkowski functionals provide a novel tool to characterize the large–scale galaxy distribution in the Universe. Here we give a brief tutorial on the basic features of these morphological measures and indicate their practical application for simulation data and galaxy redshift catalogues as examples.

## 1. Motivation

Consider the set of points in three–dimensional space supplied by galaxy coordinates of a catalogue. Let us decorate each point with a ball of radius $r$. Our task is to measure the size, shape and connectivity of the spatial pattern formed by the union set of these balls. These characteristics change with the radius $r$, which may be employed as a diagnostic parameter as illustrated in Figure 1.

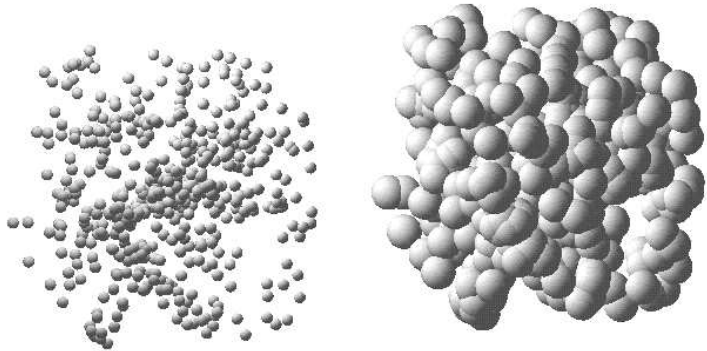

**Figure 1.** 500 points extracted from a HDM simulation decorated with balls of different radius.

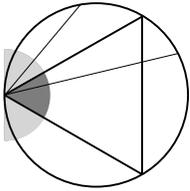

On the one hand, given an arbitrary starting point on the circle, any angle of direction within the shaded region, i.e. an angle from 0 to $\pi$, is equally probable. A longer chord lies within the dark shaded region which encompasses an angle of $\pi/3$ altogether; so this reasoning leads to a probability of $\frac{1}{3}$.

On the other hand, given an arbitrary direction, the chord takes any distance between zero and the circle's radius (shaded region) from the centre with the same probability; longer chords have distances from zero to half the radius (dark shaded region), therefore the probability is $\frac{1}{2}$.

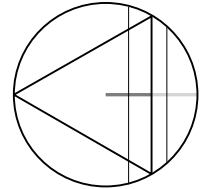

**Figure 2.** Bertrand's paradoxon: Arbitrary chords inside a circle may be longer or shorter than the side of the inscribed triangle — but what is the probability for them to be longer?

In 1889, Bertrand attempted to solve the problem in geometric probability theory shown in Figure 2 and came up with two different results. Poincaré suggested to overcome such ambiguities by tying probability measures to geometric groups. Minkowski [1] took a first systematic approach setting off the evolution of the field of mathematics now known as integral geometry.

## 2. Fundamentals

### 2.1. Hadwiger's theorem

A possible starting point for integral geometry [2] is some manifold with a group of transformations $\mathcal{G}$. Usually we are dealing with $d$–dimensional Euclidean space for which the natural choice is the group that contains as subgroups rotations and translations. One can then consider the set $\mathcal{K}$ of convex bodies embedded in this space and, as an extension, the so–called convex ring $\mathcal{R}$ of all finite unions of convex bodies. In order to characterize a body $B$ from the convex ring one looks for scalar functionals $M$ that satisfy the following requirements:

**Motion Invariance** The functional should be independent of the body's position and orientation in space,

$$M(gB) = M(B) \text{ for any } g \in \mathcal{G}, B \in \mathcal{R}.$$

**Additivity** Uniting two bodies adds their functionals, minus the functional of the intersection,

$$M(B_1 \cup B_2) = M(B_1) + M(B_2) - M(B_1 \cap B_2) \text{ for any } B_1, B_2 \in \mathcal{R}.$$

**Conditional Continuity** The functionals of convex approximations to a convex body converge to the functionals of the body†,

$$M(K_i) \to M(K) \text{ as } K_i \to K \text{ for } K, K_i \in \mathcal{K}.$$

One might think that these fairly general requirements leave a vast choice of such functionals. Surprisingly, Hadwiger's theorem states that in fact there are only $d+1$ independent such functionals if space is $d$–dimensional. In this sense the Minkowski functionals are unique and complete. To be more precise:

**Hadwiger's theorem** [3] Let $\mathcal{R}$ be the convex ring embedded in $d$–dimensional space. Then there exist $d+1$ functionals $M_\mu$, $\mu = 0 \ldots d$ on $\mathcal{R}$ such that any functional $M$ on $\mathcal{R}$ that is motion invariant, additive and conditionally continuous can be expressed as a linear combination of them:

$$M = \sum_{\mu=0}^{d} c_\mu M_\mu, \text{ with numbers } c_\mu.$$

Throughout the literature, hardly anybody uses these quantities themselves, but prefers to attach some constant factors. The four most common notations are $M_\mu$, $V_\mu$, $\overline{V}_\mu$ and $W_\mu$, defined as follows ($\omega_\mu$ is the volume of the $\mu$–dimensional unit ball):

$$V_\mu := \frac{\omega_{d-\mu}}{\omega_d} M_\mu, \quad \overline{V}_{d-\mu} := \frac{\omega_{d-\mu}}{\omega_d} \binom{d}{\mu} M_\mu,$$

$$W_\mu := \frac{\omega_\mu \omega_d}{\omega_{d-\mu}} M_\mu, \quad \text{with} \quad \omega_\mu := \frac{\pi^{\mu/2}}{\Gamma(1+d/2)}.$$

In three–dimensional Euclidean space, these functionals have a direct geometric interpretation as listed in Table 1. Apart from figures, where we will switch to the $V_\mu$–notation in order to get the easily interpretable Euler characteristic without a constant factor, we will stick to the $M_\mu$–notation throughout this paper since it offers the most convenient representation of some central formulae.

**Table 1.** The most common notations for Minkowski functionals in three–dimensional space expressed in terms of the corresponding geometric quantities.

|   | geometric quantity | $\mu$ | $M_\mu$ | $V_\mu$ | $W_\mu$ | $\overline{V}_{3-\mu}$ | $\omega_\mu$ |
|---|---|---|---|---|---|---|---|
| $V$ | volume | 0 | $V$ | $V$ | $V$ | $V$ | 1 |
| $A$ | surface | 1 | $A/8$ | $A/6$ | $A/3$ | $A/2$ | 2 |
| $H$ | mean curvature | 2 | $H/2\pi^2$ | $H/3\pi$ | $H/3$ | $H/\pi$ | $\pi$ |
| $\chi$ | Euler characteristic | 3 | $3\chi/4\pi$ | $\chi$ | $4\pi\chi/3$ | $\chi$ | $4\pi/3$ |

† This applies to convex bodies only, *not* to the whole convex ring.

Now for the geometric interpretation: The first functional equals the volume $V$ of the body, the second one is the surface area $A$. The third functional corresponds to the integral mean curvature $H$ of the body's surface and provides information about shape. Lastly, the fourth functional can be interpreted as the Euler characteristic $\chi$ which is a purely topological quantity. It is related to the genus $g$ via $\chi = 1 - g$ and can be calculated using the simple formula

$$\chi = \text{number of components} - \text{number of tunnels} + \text{number of cavities}.$$

## 2.2. The principal kinematical formula of integral geometry

One of the central issues of integral geometry is the calculation of mean values over motions of bodies. In order to do this one needs some measure $dg$ to perform integration on the group of motions $\mathcal{G}$. Demanding invariance of the measure under motions seems a sensible requirement and indeed gives a measure that is well–defined and even unique up to a multiplicative constant for most groups of interest. In order to visualize this so-called *Haar measure* it is best to stick to a certain parametrization of the group under consideration. Luckily, most applications involve calculations that can be performed within the framework of integral geometry and do not require an explicit representation of the group of motions.

Taking two bodies $A$ and $B$ from the convex ring we will fix the position and orientation of $A$ and allow $B$ to move around, i.e. apply any transformation $g$ from the group of motions $\mathcal{G}$. The resulting intersection $A \cap gB$ is again a member of the convex ring and thus has Minkowski functionals $M_\mu(A \cap gB)$. Finally, given a Haar measure $dg$ on $\mathcal{G}$, we can take the mean value over the group of motions and think of ways to calculate it from the functionals of $A$ and $B$ alone. The result is the following

**Principal kinematical formula** [3]

$$\int_\mathcal{G} dg M_\mu(A \cap gB) = \sum_{\nu=0}^{\mu} \binom{\mu}{\nu} M_\nu(A) M_{\mu-\nu}(B).$$

## 2.3. Minkowski functionals of a Poissonian distribution

The pincipal kinematical formula also offers an elegant way of obtaining explicitly the Minkowski functionals of Poisson distributed bodies [4, 5]. The Haar measure gives equal weight to any position and orientation of a body in space, so an integration over the group of motions corresponds to taking the mean value over the Poisson distribution of one body $K$. Let us distribute $N$ identical, yet independently distributed bodies $K_i$, each with Minkowski functionals $M_0 \ldots M_3$ in the region $\mathcal{V}$†. Any configuration $B_N$ can be obtained by applying a motion to each of the individual bodies, and the mean values can be calculated by multiple integration with the product measure $d\mu_N$:

$$B_N := \bigcup_{i=1}^{N} g_i K_i, \qquad d\mu_N := \prod_{i=1}^{N} \frac{dg_i}{|\mathcal{V}|} \quad \text{with} \quad \int_{\mathcal{G}^N} d\mu_N = 1.$$

† $\mathcal{V}$ has to be finite in order to ensure uniqueness of the Haar measure.

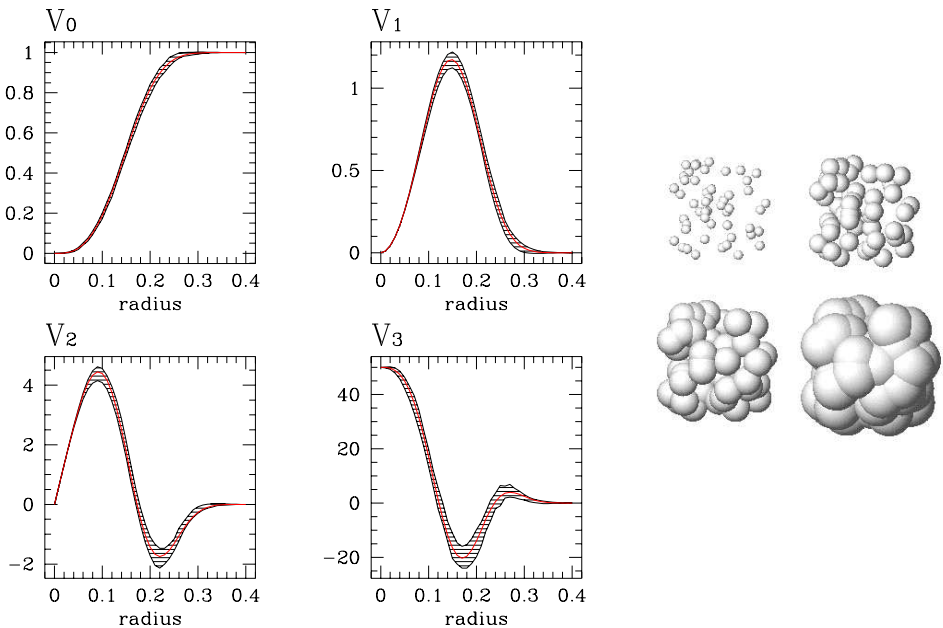

**Figure 3.** Analytical and numerical Minkowski functionals for a Poisson process plotted against the radius of balls measured in unit lengths: The shaded regions show the variance of a sample of 50 realizations of the Poisson distribution. Note that the variance is already fairly small for as few as 50 balls per unit volume. To the right, a unit box containing a typical configuration is shown at various radii displaying different topologies as explained in the text.

Using this, one can explicitly calculate the mean values $m_\mu$ of the Minkowski functionals per volume [4, 5]. Enlarging number and volume to infinity such that the density $\rho = N/|\mathcal{V}|$ remains constant one arrives at the limits:

$$\begin{aligned} m_0 &= 1 - e^{-\rho M_0}, & m_2 &= e^{-\rho M_0}(\rho M_2 - \rho^2 M_1^2), \\ m_1 &= e^{-\rho M_0} \rho M_1, & m_3 &= e^{-\rho M_0}(\rho M_3 - 3\rho^2 M_1 M_2 + \rho^3 M_1^3). \end{aligned}$$

The values are plotted, together with numerical results, in Figure 3 for a density of 50 balls per unit volume. One can see that, as the radius increases, the volume (functional $V_0$) is filled until reaching complete coverage at density 1, while the topology undergoes a number of changes which can be seen in the Euler characteristic (functional $V_3$) as well as in the configurations shown on the right. Very tiny balls (top left, radius .05) remain isolated from each other and so the Euler characteristic $\chi$ is close to the number of balls for the lowest radii. As the radius increases, balls join (top right, radius .1) and hence $\chi$ decreases, even below zero as more and more tunnels are formed in the configuration (bottom left, radius .15). This behaviour reaches a turning point when these tunnels are blocked to form closed cavities (bottom right, radius .25) which give again a positive contribution to the Euler characteristic. Finally, these cavities are closed one by one and so we end up with all space filled, and $\chi$ equal to zero.

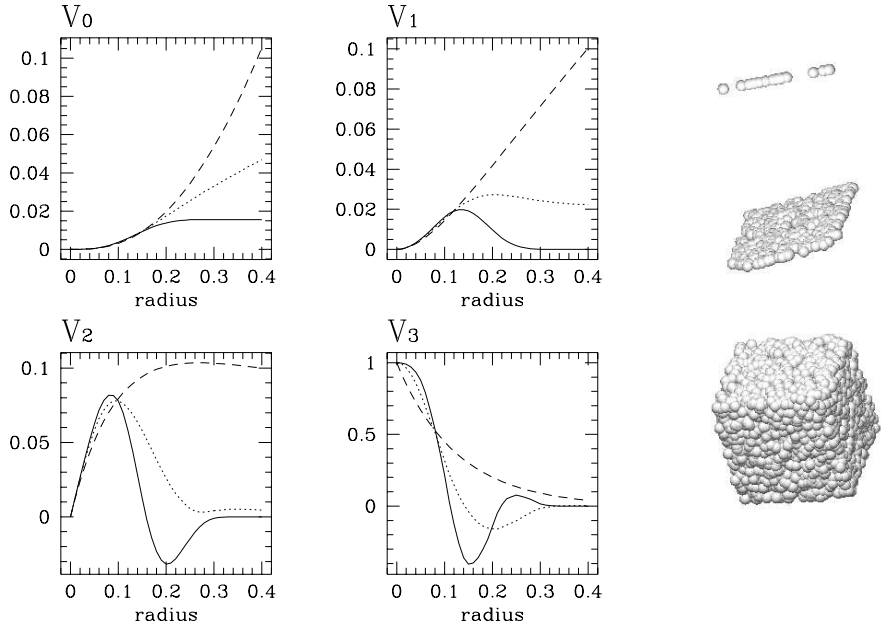

**Figure 4.** Minkowski functionals for idealized structures are highly efficient discriminators. To the left, Minkowski functionals per point and unit volume are shown for a Poisson distribution on a line (dashed), on a plane (dotted) and inside the whole space (solid). In all distributions the points were seperated by the same mean distance of .25 unit lengths; in the example configurations to the right the points have a mean seperation of .05 unit lengths and are decorated with balls of the same radius.

## 2.4. Clusters, walls and filaments

The principal kinematical formula also allows some analytical calculations for point processes beyond the simple Poissonian case. Since Minkowski functionals have already yielded some promising preliminary results [6] concerning discrimination of filamentary and wall–like structures in the Universe, we have calculated analytical values for the Minkowski functionals of idealized filaments and walls, i.e. points on an infinitely extended straight line distributed according to a Poisson distribution (on a flat plane, respectively). The results for the three distributions are shown in Figure 4 for equal mean distance of the points.

## 3. Applications

Examples of the discriminatory power of the Minkowski functionals are shown in Mecke, Buchert and Wagner [7], Buchert [8], and Platzöder and Buchert [6]. In the following we discuss the application of Minkowski functionals to simulation data and to redshift catalogues.

### 3.1. Simulations

As an example for the analysis of simulation data we calculated the Minkowski functionals for a HDM simulation (using first–order Lagrangian perturbation theory), with $h = 0.5$, $\Omega = 1$, and primordial spectral index $n = +1$ in a cubic box with side length 200 Mpc using $64^3$ particles. We extracted randomly 5 different subsamples with 5000 points. In Figure 5 we show the mean and the standard error of the Minkowski functionals calculated in a periodic box. Although using only five subsamples the errors are very small. In this sense the Minkowski functionals are *robust*. Using this approach we are able to compare simulations. A comparison with data derived from catalogues is only possible with a suitable galaxy or cluster extraction scheme and a treatment of the boundaries (see below). Since Minkowski functionals depend on the number density, comparisons are only possible for samples of equal number density. Currently only scaling relations for the Poisson process are known to us.

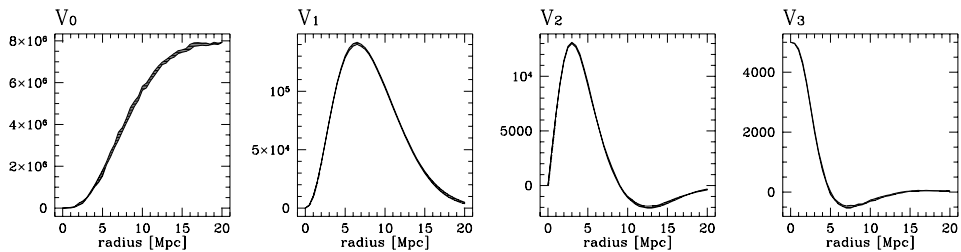

**Figure 5.** Minkowski functionals of a HDM simulation.

### 3.2. Redshift catalogues

Integral geometry offers a concise way of dealing with boundaries [4]. Let $D$ be the window (the sample geometry) through which we look at $N$ galaxies. $B_r = \bigcup_{i=0}^{N} B_r(i)$ is the union of all balls $B_r(i)$ of radius $r$ centered on the $i$–th galaxy respectively. One has to calculate the Minkowski functionals of the intersection of the union of all balls with the window, $M_\mu(B_r \cap D)$ and the Minkowski functionals of the window itself $M_\mu(D)$. Therefore, we not only have to compute the intersection of two or three balls [7], but we also have to compute the intersections of a ball with the boundary of $D$.

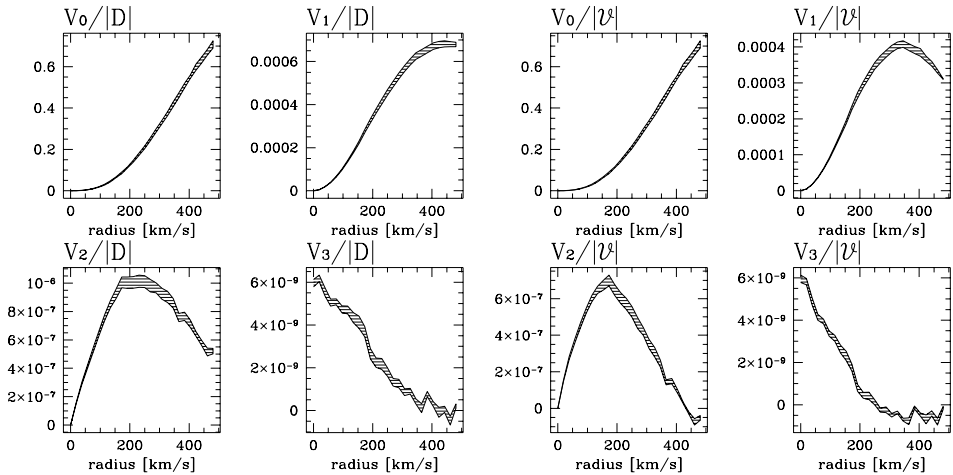

**Figure 6.** To the left the densities of the Minkowski functionals $V_\mu(B_r \cap D)/|D|$ within the sample window $D$ are depicted, to the right the densities of the Minkowski functionals $V_\mu(B_r)/|\mathcal{V}|$ of the Universe. As expected, the volume density is the same for both sets, while the other functionals clearly exhibit the removal of boundary effects. The errors are bootstrap errors, we extracted ten samples with 143 galaxies from the 153 galaxies in the cube. Again the errors are small, we nearly gain nothing in extracting fifty bootstrap samples instead of ten.

The quantities $M_\mu(B_r \cap D)$ are well suited for the analysis of redshift catalogues since we do not assume anything on the point distribution and still analyze all galaxies in the sample. Using the same window $D$ one is able to compare different catalogues and investigate their homogeneity and isotropy. It is not necessary to impose artificial assumptions on the underlying point process like using periodic boundary conditions (in the case of a cubic sample) or any other way of embedding (e.g. Poisson).

Although we argued in favour of *not* imposing assumptions on the underlying point process we want to show another way of extracting information from the catalogue, assuming homogeneity and isotropy. Then we get for the average under movements of the window

$$\langle M_\mu(B_r \cap D) \rangle = \frac{1}{|\mathcal{V}|} \int_{\mathcal{G}} M_\mu(B_r \cap gD) \mathrm{d}g.$$

$\mathcal{V}$ again is a region significantly larger than $|D|$ (e.g., the whole Universe). If we additionally assume that within $D$ we have a fair sample of the Universe, the averages greatly simplify:

$$\langle M_\mu(B_r \cap D) \rangle = M_\mu(B_r \cap D).$$

Using the principal kinematical formula we then conclude

$$M_\mu(B_r \cap D) = \frac{1}{|\mathcal{V}|} \sum_{\nu=0}^{\mu} \binom{\mu}{\nu} M_\nu(B_r) M_{\mu-\nu}(D).$$

With these *strong* assumptions and using the computed values of $M_\mu(B_r \cap D)$ and $M_\mu(D)$, we are able to extract the densities of the Minkowski functionals $M_\mu(B_r)/|\mathcal{V}|$, thus removing the contributions of the boundary. We get†

$$\frac{M_\mu(B_r)}{|\mathcal{V}|} = \frac{M_\mu(B_r \cap D)}{M_0(D)} - \sum_{\nu=0}^{\mu-1} \binom{\mu}{\nu} \frac{M_\nu(B_r)}{|\mathcal{V}|} \frac{M_{\mu-\nu}(D)}{M_0(D)}.$$

These expressions are more suitable for computational purposes than the explicit solution of the system.

To illustrate this procedure we calculate the density of the Minkowski functionals in the Universe based on the intersection of the Universe with a cube selected from the CfA catalog as discussed by Gott et al. [9], with side length $s_0 = 5000/\sqrt{3} h^{-1} \text{kms}^{-1}$. We put a cube, our window $D$, with sidelength $s = s_0/2$ into the center of the sample cube. This is necessary since we calculate the Minkowski functionals for balls up to a radius $s_0/6$. For the window we get

$$V_0(D) = |D| = s^3, \quad V_1(D) = s^2, \quad V_2(D) = s, \quad V_3(D) = 1.$$

In Figure 6 we show the densities of the Minkowski functionals $V_\mu(B_r \cap D)/|D|$ including the contributions of the intersections with the window $D$, and the reconstructed densities of the Minkowski functionals $V_\mu(B_r)/|\mathcal{V}|$ of the Universe i.e., the Minkowski functionals after removing the contributions of the window $D$. Again we want to emphasize that *no* assumptions enter in calculating $V_\mu(B_r \cap D)/|D|$, but the reconstruction of $V_\mu(B_r)/|\mathcal{V}|$ is relying on homogeneity and isotropy.

This is meant as an example; this cube is certainly not a fair sample.

## 4. The software and its future

The ANSI C code for computing the Minkowski functionals of balls in a box with periodic boundary conditions may be obtained via e–mail from buchert@stat.physik.uni-muenchen.de. The algorithm follows the description in Mecke, Buchert, and Wagner [7] and calculates the $V_\mu$ measures. Memory requirements are moderate (less than 5 MByte for samples of up to 10,000 points), while CPU time strongly depends on the number of points and on the clustering properties of the point distribution. A typical example is a Poisson distribution of 1,000 points, which took approximately 40 minutes on a HP 715/80 workstation. A parallelized version using pvm3 and a version dealing with boundaries in the described way are in the test stage. Additionally we are working on a program for structure discrimination using the analytical formulae for filaments, walls, and clusters.

---

† We use the convention $\sum_{i=0}^{-1} a_i = 0$.

## Acknowledgments

For valuable discussions we are deeply indebted towards Herbert Wagner. We thank Vicent Martínez for supplying the CfA cube and Arno G. Weiss for the HDM simulation data. During the summer school we appreciated encouraging discussions with Joel Primack, Peter Coles, and other lecturers and students. Martin Kerscher and Thomas Buchert acknowledge financial support by the "Sonderforschungsbereich 375–95 für Astro–Teilchenphysik" der Deutschen Forschungsgemeinschaft. Jens Schmalzing thanks the Stiftung Maximilianeum and, together with Martin Kerscher, the Italian Physical Society for financial support during the summer school in Varenna.

## References

[1] Minkowski H. 1903 Volumen und Oberfläche *Mathematische Annalen* **57** (Leipzig: B. G. Teubner) 447-495
[2] Santaló L. A. 1976 *Integral Geometry and Geometric Probability* (Reading, MA: Addison–Wesley)
[3] Hadwiger H. 1957 *Vorlesungen über Inhalt, Oberfläche und Isoperimetrie* (Berlin: Springer).
[4] Mecke K. and Wagner H. 1991 Euler characteristic and related measures for random geometric sets *J. Stat. Phys.* **64** 843.
[5] Mecke K. 1994 *Integralgeometrie in der Statistischen Physik: Perkolation, komplexe Flüssigkeiten und die Struktur des Universums* (Thun, Frankfurt/Main: Deutsch).
[6] Platzöder M., Buchert T. 1995 Applications of Minkowski functionals for the statistical analysis of dark matter models $1^{\text{st}}$ *SFB workshop on Astro-particle physics*, in press
[7] Mecke K., Buchert T. and Wagner H. 1994 Robust morphological measures for large–scale structure in the Universe *Astron. Astrophys.* **288** 697.
[8] Buchert T. 1995 Robust morphological measures for large–scale structure *11th Potsdam Workshop on Large-Scale Structure in the Universe (Geltow)* eds Mücket J., Gottlöber S. and Müller, V. (World Scientific) in press.
[9] Gott J.R., Melott A.L. and Dickinson M. 1986 The sponge–like topology of large–scale structure in the Universe *Ap. J.* **306** 341.